\documentclass[showpacs,prl,twocolumn,aps]{revtex4}

\usepackage{bm}
\usepackage{amsmath}
\usepackage{amssymb}
\usepackage{latexsym} 
\usepackage{amsfonts} 
\usepackage{epsfig}
\usepackage{psfrag} 

\newcommand{\ket}[1]{\left|#1\right>} 
\newcommand{\bra}[1]{\left<#1\right|}

\newcommand{\f}[1]{\mbox{\boldmath$#1$}}

\newcommand{\bea}{\begin{eqnarray}}
\newcommand{\ea}{\end{eqnarray}}
\newcommand{\eea}{\end{eqnarray}}
\newcommand{\ord}{{\cal O}}

\begin{document}
\title{Quantum simulator for the $\f{O(3)}$ nonlinear sigma model}
\author{Ralf Sch\"utzhold and Sarah Mostame}
\affiliation{Institut f\"ur Theoretische Physik, 
Technische Universit\"at Dresden, D-01062 Dresden, Germany\\
Email: {\tt schuetz@theory.phy.tu-dresden.de; sarah@theory.phy.tu-dresden.de}}
\begin{abstract} 
We propose a design for the construction of a laboratory system based 
on present-day technology which reproduces and thereby simulates the
quantum dynamics of the $O(3)$ nonlinear sigma model.
Apart from its relevance in condensed-matter theory, this strongly
interacting quantum field theory serves as an important toy model for
quantum chromo-dynamics (QCD) since it reproduces many crucial
properties of QCD.   
\end{abstract} 
  
\maketitle
 
%%%%%%%%%%%%%%%%%%%%%%%%%%%%%%%%%%%%%%%%%%%%%%%%%%%%%%%%%%%%%%%%%%%%%%%%%%%%%%%
%%%%%%%%%%%%%%%%%%%%%%%%%%%%%%%%%%%%%%%%%%%%%%%%%%%%%%%%%%%%%%%%%%%%%%%%%%%%%%%
{\em Introduction}\quad
%%%%%%%%%%%%%%%%%%%%%%%%%%%%%%%%%%%%%%%%%%%%%%%%%%%%%%%%%%%%%%%%%%%%%%%%%%%%%%%
%%%%%%%%%%%%%%%%%%%%%%%%%%%%%%%%%%%%%%%%%%%%%%%%%%%%%%%%%%%%%%%%%%%%%%%%%%%%%%%
%
In many areas of physics, progress has been thwarted by our lack of
understanding strongly interacting quantum systems with many degrees
of freedom such as quantum field theories.
Beyond perturbation theory with respect to some parameter or
semiclassical models/methods, there are not many analytical tools
available for the treatment of these systems.
Numerical methods are hampered by the exponentially increasing amount
of resources required for the simulation of quantum systems with many 
degrees of freedom in general.
However, this obstacle applies to classical computers only -- quantum
computers will be able to simulate other quantum systems with
polynomial effort \cite{quasi}.
But as long as universal quantum computers of sufficient size 
(e.g., number of QuBits) are not available, one has to search for
alternatives. 
One possibility is to design a special quantum system in the
laboratory which reproduces the Hamiltonian of a particular quantum
field theory of interest.
This designed quantum system can then be regarded as a special quantum
computer (instead of a universal one) which just performs the desired
quantum simulation.
In this Letter, we propose such a quantum simulator for the example of
the $O(3)$ nonlinear sigma model and demonstrate that it can be
constructed using present-day technology.

%%%%%%%%%%%%%%%%%%%%%%%%%%%%%%%%%%%%%%%%%%%%%%%%%%%%%%%%%%%%%%%%%%%%%%%%%%%%%%%
%%%%%%%%%%%%%%%%%%%%%%%%%%%%%%%%%%%%%%%%%%%%%%%%%%%%%%%%%%%%%%%%%%%%%%%%%%%%%%%
{\em The Model}\quad
%%%%%%%%%%%%%%%%%%%%%%%%%%%%%%%%%%%%%%%%%%%%%%%%%%%%%%%%%%%%%%%%%%%%%%%%%%%%%%%
%%%%%%%%%%%%%%%%%%%%%%%%%%%%%%%%%%%%%%%%%%%%%%%%%%%%%%%%%%%%%%%%%%%%%%%%%%%%%%%
%
The 1+1 dimensional $O(N)$ $\sigma$-model
\cite{origin,asy-free,instantons,general,exact,lattice,renormalize} is
described by the $O(N)$ and Poincar\'e invariant action  
\bea
\label{lag-sigma}
{\cal L}
=
\frac{\hbar}{2c}\,
\partial_\nu\f{\sigma}\cdot\partial^\nu\f{\sigma}
=
\frac{\hbar}{2c}
\left[(\partial_t\f{\sigma})^2-c^2(\partial_x\f{\sigma})^2\right]
\,,
\ea
with the internal vector 
$\f{\sigma}=(\sigma_1,\sigma_2,\dots,\sigma_N)\in{\mathbb R}^N$
reflecting the $O(N)$-symmetry.
So far, this theory describes $N$ independent free fields, but the
constraint 
\bea
\label{norm-sigma}
\f{\sigma}^2
=
\sigma_1^2+\sigma_2^2+\dots+\sigma_N^2
=
\frac{N}{g^2}
\,,
\ea
introduces an interaction corresponding to the coupling $g>0$. 
For vanishing coupling $g\downarrow0$, the curvature of the constraint
sphere ($\f{\sigma}^2=N/g^2$) vanishes and we reproduce (locally) an
effectively free theory. 
For finite coupling $g>0$, we obtain a non-trivially interacting
theory as long as $N\geq3$, i.e., in the non-Abelian case.
The classical ground state $\f{\sigma}=\rm const$ is
$O(N)$-degenerate, but quantum interaction lifts that degeneracy and
gives the classical Goldstone modes a mass gap, 
see, e.g., \cite{exact,general}.

%%%%%%%%%%%%%%%%%%%%%%%%%%%%%%%%%%%%%%%%%%%%%%%%%%%%%%%%%%%%%%%%%%%%%%%%%%%%%%%
%%%%%%%%%%%%%%%%%%%%%%%%%%%%%%%%%%%%%%%%%%%%%%%%%%%%%%%%%%%%%%%%%%%%%%%%%%%%%%%
{\em Properties}\quad
%%%%%%%%%%%%%%%%%%%%%%%%%%%%%%%%%%%%%%%%%%%%%%%%%%%%%%%%%%%%%%%%%%%%%%%%%%%%%%%
%%%%%%%%%%%%%%%%%%%%%%%%%%%%%%%%%%%%%%%%%%%%%%%%%%%%%%%%%%%%%%%%%%%%%%%%%%%%%%%
%
Apart from its relevance in condensed-matter theory 
(partly due to its relation to spin-systems such as anti-ferromagnets, 
see, e.g., \cite{cond-mat}), 
it can be shown that the 1+1 dimensional $O(N)$ $\sigma$-model
reproduces many crucial properties of quantum chromo-dynamics (QCD)
and hence serves as an important toy model, see, e.g., \cite{general}.
The $\sigma$-model is renormalizable 
(in 1+1 dimensions, cf.~\cite{renormalize}) and
its running coupling $g(p^2)$ generates asymptotic freedom 
$g^2(p^2\gg\Lambda^2)\propto1/\ln(p^2/\Lambda^2)$, cf.~\cite{asy-free}.
In analogy to QCD, the classical scale invariance 
$x^\nu\to\Omega\,x^\nu$ is broken dynamically corresponding to the
dimensional transmutation $g\to\Lambda$. 
Furthermore, the $\sigma$-model generates non-vanishing vacuum
condensates such as $\langle\hat{\cal L}\rangle\neq0$ in the operator
product expansion and reproduces the trace anomaly 
$\langle\hat T^\nu_\nu\rangle\neq0$ (see, e.g., \cite{general}). 
It also serves as a toy model for the study of the low-energy theorems
and sum rules (see, e.g., \cite{general}). 
For $N=3$, the $\sigma$-model exhibits instantons 
(mapping of ${\mathbb S}_2$ onto ${\mathbb R}^2$, 
cf.~\cite{instantons,general}).
Furthermore, the model is exactly solvable in the large-$N$ limit,
where it corresponds to massive free fields with sub-leading 
(in $1/N$) interaction terms, see, e.g., \cite{general}.
Its complex version, the $CP(N-1)$-model, reproduces confinement
(though this is not such a striking feature in 1+1 dimensions) and the
$U(1)$-problem of QCD (see, e.g., \cite{general}). 

\begin{figure}[ht]
\centerline{\mbox{\epsfxsize=8.5cm\epsffile{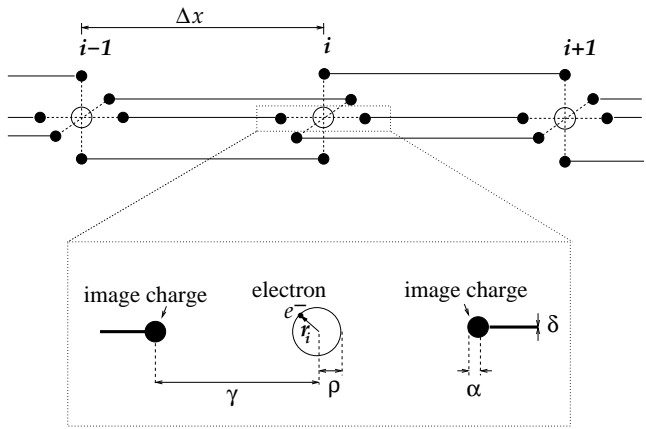}}}
\caption{Sketch of the proposed analogue quantum simulator.\\ 
The solid lines and spheres denote (super) conductors and the hollow
spheres are insulators containing single electrons.
Shown are just three elements of a long chain (top) and
a close-up view (bottom) with the involved length scales.}
\label{analogue}
\end{figure}

%%%%%%%%%%%%%%%%%%%%%%%%%%%%%%%%%%%%%%%%%%%%%%%%%%%%%%%%%%%%%%%%%%%%%%%%%%%%%%%
%%%%%%%%%%%%%%%%%%%%%%%%%%%%%%%%%%%%%%%%%%%%%%%%%%%%%%%%%%%%%%%%%%%%%%%%%%%%%%%
{\em The Analogue}\quad
%%%%%%%%%%%%%%%%%%%%%%%%%%%%%%%%%%%%%%%%%%%%%%%%%%%%%%%%%%%%%%%%%%%%%%%%%%%%%%%
%%%%%%%%%%%%%%%%%%%%%%%%%%%%%%%%%%%%%%%%%%%%%%%%%%%%%%%%%%%%%%%%%%%%%%%%%%%%%%%
%
In order to reproduce the quantum dynamics of the the 1+1 dimensional
$O(N)$ $\sigma$-model according to Eqs.~(\ref{lag-sigma}) and
(\ref{norm-sigma}), let us consider a large number of perfectly
insulating thin hollow spheres with the radius $\rho$ lined up at
equal distances $\Delta x$ with single electrons being captured by the
polarizability (inducing a finite extraction energy) on each of the
hollow spheres. 
These insulating spheres are surrounded by an arrangement of
superconducting spheres (radius $\alpha$) and wires (radius $\delta$)
as depicted in Fig.~\ref{analogue}, which generate controlled
interactions of the confined electrons via their image charges.
The involved length scales including the typical wavelength of the  
excitations $\lambda$, the distance of elements (lattice spacing) 
$\Delta x$, the distance between the insulating and the conducting
spheres $\gamma$, the radii of the insulating and conducting spheres
$\rho$ and $\alpha$ and wires $\delta$ (cf.~Fig.~\ref{analogue}) are
supposed to obey the following hierarchy 
\bea
\label{assumption}
\lambda
\gg
\Delta x
\gg
\gamma
\gg
\rho,\alpha
\gg
\delta
\,.
\ea

The total Lagrangian for the system of electrons reads
\bea
\label{lag-origin}
{\mathfrak L}=
\sum\limits_i\left[\frac{m}{2}\,\f{\dot r}_i^2-
V(\f{r}_{i+1},\f{r}_{i})\right]
\,,
\ea
with $m$ being the mass of the electrons and
$V(\f{r}_{i+1},\f{r}_{i})$ their interaction potential, where only
nearest neighbors are taken into account in view of the assumptions
(\ref{assumption}). 
In this limit, the interaction potential induced by the images of the 
electron charges $e$ simplifies to 
\bea
\label{interaction}
V(\f{r}_{i+1},\f{r}_{i})=\frac{e^2\alpha^2}{4\pi\varepsilon_0\gamma^4}\,
\frac{(\f{r}_{i+1}-\f{r}_{i})^2}{4\alpha+\Delta x/\ln(\Delta x/\delta)}
\,,
\ea
where the first addend in the denominator on the right-hand side is
due to the capacitance of the conducting spheres
$4\pi\varepsilon_0\alpha$ and the second one due to the capacitance of
the long wires $2\pi\varepsilon_0\Delta x/\ln(\Delta x/\delta)$.
Comparing the resulting Lagrangian in Eqs.~(\ref{lag-origin}) and
(\ref{interaction}) with the one in Eq.~(\ref{lag-sigma}), we can read
off the effective propagation speed 
\bea
\label{c-eff}
c_{\rm eff}=c_0\sqrt{\frac{e^2}{4\pi\varepsilon_0mc_0^2}\,
\frac{2\alpha^2\Delta x^2/\gamma^4}{4\alpha+\Delta x/\ln(\Delta x/\delta)}}
\,.
\ea
Since the first term under the root represents the classical electron
radius (of order $10^{-15}$ m), the effective propagation speed 
$c_{\rm eff}$ is much smaller than the speed of light in vacuum 
$c_0 \gg c_{\rm eff}$ for realistic parameters (see below), 
i.e., we obtain a large slow-down.  
Furthermore, we may identify the effective coupling for $N=3$
\bea
\label{g-eff}
g_{\rm eff}
=
\sqrt{3}\,
\frac{\gamma}{\rho}\,
\sqrt[4]{
\frac{4\pi\varepsilon_0\hbar^2}{me^2}\,
\frac{4\alpha+\Delta x/\ln(\Delta x/\delta)}{2\alpha^2}
}
\,,
\ea
where the first term under the root is the classical electron radius
over the square of the fine structure constant.
The value of the effective coupling can be tuned by varying the ratio
$\gamma/\rho\gg1$ and may well be of order one (see parameters below).
Strictly speaking, the above equation determines the value of the
running coupling $g_{\rm eff}(p^2)$ at a length scale corresponding to
the lattice spacing $\Delta x$ (lattice renormalization scheme). 
In complete analogy to $\Lambda_{\rm QCD}$,  the coupling 
$g_{\rm eff}^2(p^2\gg\Lambda_\sigma^2)\propto1/\ln(p^2/\Lambda^2_\sigma)$
determines the induced scale of dynamical symmetry breakdown
$\Lambda_\sigma$ of the $\sigma$-model (dimensional transmutation).
This important quantity sets all other length scales such as the mass
gap (see, e.g., \cite{exact}) and must satisfy the condition
(\ref{assumption}) for consistency, i.e., $\Lambda_\sigma\Delta x\ll1$.
Finally, identifying (again for $N=3$)
\bea
\label{sigma-eff}
\f{\sigma}(x=i\Delta x)=
\frac{\sqrt{3}}{g_{\rm eff}}\,\frac{\f{r}_{i}}{\rho}
\,,
\ea
the continuum limit 
($\sum_i\Delta x\to\int dx$ for $\lambda\gg\Delta x$)
of Eq.~(\ref{lag-origin}) generates the Lagrangian (\ref{lag-sigma})
of the $O(3)$ nonlinear sigma model with the constraint
(\ref{norm-sigma}) being implemented by $\f{r}_{i}^2=\rho^2$. 

%%%%%%%%%%%%%%%%%%%%%%%%%%%%%%%%%%%%%%%%%%%%%%%%%%%%%%%%%%%%%%%%%%%%%%%%%%%%%%%
%%%%%%%%%%%%%%%%%%%%%%%%%%%%%%%%%%%%%%%%%%%%%%%%%%%%%%%%%%%%%%%%%%%%%%%%%%%%%%%
{\em Disturbances}\quad
%%%%%%%%%%%%%%%%%%%%%%%%%%%%%%%%%%%%%%%%%%%%%%%%%%%%%%%%%%%%%%%%%%%%%%%%%%%%%%%
%%%%%%%%%%%%%%%%%%%%%%%%%%%%%%%%%%%%%%%%%%%%%%%%%%%%%%%%%%%%%%%%%%%%%%%%%%%%%%%
%
Of course, for a realistic proposal, it is essential to estimate the
impact of the contributions which have been omitted so far.
The additional kinetic terms due to inductances $L$ of the wires are
negligible $LI^2 \ll m\f{\dot r}^2$ provided that 
\bea
\label{inductance}
4\frac{\alpha}{\Delta x}\left(\frac{c_{\rm eff}}{c_0}\right)^2
\ln\left(\frac{\Delta x}{\delta}\right)
\ll1
\ea
holds, i.e., for a sufficiently large slow-down (as one would expect). 
For the same reason, the influence of the zero-point fluctuations of
the electromagnetic field (inductance of free space) is negligible.

In contrast to sequential quantum algorithms, where errors may
accumulate over many operations, the quantum simulation under
consideration is basically a ground state problem and hence more
similar to adiabatic quantum computing \cite{adiabatic}.
In this case, decoherence can be neglected as long as the interaction
energies of the disturbances are much smaller than the energy gap
between the ground state and the first excited state \cite{adiabatic}.
For the nonlinear $\sigma$-model, this gap is determined by the
induced scale $\Lambda_\sigma$ (in analogy to QCD).
Therefore, the energies of all perturbations 
(e.g., impurities in the material) must be much smaller than
the gap of order $\hbar c_{\rm eff} \Lambda_\sigma$.
In particular, in order to see quantum behavior 
(where the Heisenberg uncertainty relation becomes important), the
temperature must be small enough
\bea
\label{temperature}
k_{\rm B}T \ll \hbar c_{\rm eff} \Lambda_\sigma
\,.
\ea
Another issue concerns the spins of the electrons, which have been
omitted so far. 
Fortunately, we may fix the electron spins by a small external
magnetic field (see the next paragraph) and the various spin-spin and
especially spin-orbit coupling terms are negligible 
(in comparison to $\hbar c_{\rm eff} \Lambda_\sigma$) 
for the parameters provided below.  

%%%%%%%%%%%%%%%%%%%%%%%%%%%%%%%%%%%%%%%%%%%%%%%%%%%%%%%%%%%%%%%%%%%%%%%%%%%%%%%
%%%%%%%%%%%%%%%%%%%%%%%%%%%%%%%%%%%%%%%%%%%%%%%%%%%%%%%%%%%%%%%%%%%%%%%%%%%%%%%
{\em Phase Diagram}\quad
%%%%%%%%%%%%%%%%%%%%%%%%%%%%%%%%%%%%%%%%%%%%%%%%%%%%%%%%%%%%%%%%%%%%%%%%%%%%%%%
%%%%%%%%%%%%%%%%%%%%%%%%%%%%%%%%%%%%%%%%%%%%%%%%%%%%%%%%%%%%%%%%%%%%%%%%%%%%%%%
%
Before investigating the impact of an external magnetic field, let us
turn to the phase diagram of the nonlinear $\sigma$-model in terms of
the temperature $T$ and the chemical potential $\mu$.
For low temperatures $k_{\rm B}T \ll \hbar c_{\rm eff} \Lambda_\sigma$
and small chemical potentials $\mu\ll \hbar c_{\rm eff} \Lambda_\sigma$, 
we basically get the usual vacuum state.
Note that the introduction of a chemical potential necessitates the
definition of a particle number (which is a nontrivial issue in 
interacting theories). 
In the $\sigma$-model, this can be achieved by means of the Noether
current corresponding to the global $O(3)$ invariance 
$\f{j}_\nu=\f{\sigma}\times\partial_\nu\f{\sigma}$
and the associated global charge along some internal axis $\f{n}$ with
$\f{n}^2=1$ 
\bea
\label{charge}
Q=\frac{1}{c_{\rm eff}}\,\f{n}\cdot\int dx\;\f{\sigma}\times\f{\dot\sigma}
\,.
\ea
For the laboratory system, the Noether charge $Q$ is just the
total (orbital) angular momentum in units of $\hbar$. 
Note that still many charges $Q\gg1$ are required to generate one
magnetic flux quantum (due to $c_0 \gg c_{\rm eff}$). 

In terms of the chemical potential defined with respect to this
(dimensionless) Noether charge, the grand-canonical Hamiltonian $\hat
H_{\rm gc}$ reads 
\bea
\label{grand}
\hat H_{\rm gc}=\hat H_0+\mu_N\hat N=\hat H_0+\mu_Q\hat Q
\,.
\ea
Translating this expression back to our laboratory system in
Eq.~(\ref{lag-origin}), we observe that the chemical potential exactly
corresponds to an external magnetic field $\f{B}$ inducing the
additional term  $\f{\dot r}\cdot\f{A}=
\f{\dot r}\cdot(\f{r}\times\f{B})/3=\f{B}\cdot(\f{\dot r}\times\f{r})/3$ 
\bea
\label{chemical}
\mu_{\rm eff}=\frac{e\hbar}{3m}\,B
\,.
\ea
(The second-order term $e^2\f{A}^2/m$ is three orders of magnitude
smaller for the parameters given below and can be neglected.)
When the effective chemical potential $\mu_{\rm eff}$ exceeds the
energy gap of order $\hbar c_{\rm eff} \Lambda_\sigma$, the structure
of the ground state changes and the above Noether current $\f{j}_\nu$
acquires a non-vanishing expectation value 
(quantum phase transition, see, e.g., \cite{exact}).
At the critical field
$B_{\rm crit}={\cal O}(mc_{\rm eff}\Lambda_\sigma/e)$ 
where this quantum phase transition occurs, the energy of the electron 
spins is of the same order as the gap 
$\f{\mu}_s\cdot\f{B}={\cal O}(\hbar c_{\rm eff} \Lambda_\sigma)$ 
and thus much bigger than the temperature.
Hence one can fix the electron spins with much smaller external
magnetic fields $B \ll B_{\rm crit}$ without disturbing the vacuum
state too much.
On the other hand, it is also possible to explore the full phase
diagram 
(e.g., cross the quantum phase transition, monitored by a SQUID)
by increasing the external magnetic field -- which is
completely equivalent to changing the chemical potential 
(and hence the number of particles).
For the set of parameters discussed below, the critical field 
$B_{\rm crit}$ is of order milli-Tesla.

%%%%%%%%%%%%%%%%%%%%%%%%%%%%%%%%%%%%%%%%%%%%%%%%%%%%%%%%%%%%%%%%%%%%%%%%%%%%%%%
%%%%%%%%%%%%%%%%%%%%%%%%%%%%%%%%%%%%%%%%%%%%%%%%%%%%%%%%%%%%%%%%%%%%%%%%%%%%%%%
{\em Experimental Parameters}\quad
%%%%%%%%%%%%%%%%%%%%%%%%%%%%%%%%%%%%%%%%%%%%%%%%%%%%%%%%%%%%%%%%%%%%%%%%%%%%%%%
%%%%%%%%%%%%%%%%%%%%%%%%%%%%%%%%%%%%%%%%%%%%%%%%%%%%%%%%%%%%%%%%%%%%%%%%%%%%%%%
%
The aforementioned constraints, in particular Eqs.~(\ref{assumption})
and (\ref{temperature}), provide the frame of a window of opportunity
for the experimental realization of the proposed quantum simulator --
which is (fortunately) open to present-day technology.
Let us first explore the limit set by the ultra-low temperatures.
For solid bodies of reasonable size, one can reach temperatures of
order 10 $\mu$K by electron gas cooling via spin relaxation.
If we choose our parameters according to
$\delta=100\;\rm nm$,
$\rho=400\;\rm nm$,
$\alpha=500\;\rm nm$,
$\gamma=2.5\;\mu\rm m$, and 
$\Delta x=12.5\;\mu\rm m$,
we obtain 
$g_{\rm eff}=\ord(1)$,
$c_{\rm eff}\approx 10^4\,\rm m/s$,
$\Lambda_\sigma^{-1}\approx 125\;\mu\rm m$,
and $\hbar c_{\rm eff} \Lambda_\sigma$ corresponds to 600 $\mu$K,
which satisfies all of the above assumptions reasonably well. 
Alternatively, we may start from the present state of nanotechnology
which facilitates the production of nanowires with a radius of order
nanometer.
If we explore this limit and choose 
$\delta=1\;\rm nm$,
$\rho=12\;\rm nm$,
$\alpha=5\;\rm nm$,
$\gamma=25\;\rm nm$, and 
$\Delta x=125\;\rm nm$,
we obtain a similar value for $g_{\rm eff}$ and 
$c_{\rm eff}\approx 10^5\,\rm m/s$, but now 
$\hbar c_{\rm eff} \Lambda_\sigma$ 
corresponds to a temperature of order Kelvin. 
The range between $\mu$K and fractions of a Kelvin as well
as between nanometers and micrometers provides a two or three orders
of magnitude wide window of opportunity and the optimum experimental
parameters are probably somewhere in the middle.

The thin superconducting wires can be switched on and off by local
variations of the temperature (below and above the critical value).
If the interaction $V(\f{r}_{i+1},\f{r}_{i})$ is switched off, the
energy spectrum of the electrons is determined by the usual 
spherical harmonics 
\bea
\label{ell}
E_\ell=\frac{\hbar^2}{2m}\,\frac{\ell(\ell+1)}{\rho^2}
\,.
\ea
The energy gap between the s-state $\ell=0$ and the p-state $\ell=1$,
i.e., without interaction $V(\f{r}_{i+1},\f{r}_{i})$, is one order of
magnitude larger than with interaction $\hbar c_{\rm eff} \Lambda_\sigma$. 
Consistently, the interaction potential $V(\f{r}_{i+1},\f{r}_{i})$
between the electrons on different spheres is of the same order of
magnitude as the gap between the s-state $\ell=0$ and the p-state
$\ell=1$ on a single sphere leading to strong entanglement of the
ground state. 
If we want to switch on the interaction $V(\f{r}_{i+1},\f{r}_{i})$ 
adiabatically (e.g., via changing the temperature of the wires)
satisfying the condition for the adiabatic theorem 
$|\bra{\psi_0}d\hat H(t)/dt\ket{\psi_1}|/(\Delta E_{01})^2\ll1$
in order to stay in the ground state $\ket{\psi_0}$, the typical
adiabatic switching time should be longer than a few picoseconds.
Finally, for the parameters discussed above, the various spin-spin and  
spin-orbit coupling energies are at least two orders of magnitude
smaller than $\hbar c_{\rm eff} \Lambda_\sigma$.

%%%%%%%%%%%%%%%%%%%%%%%%%%%%%%%%%%%%%%%%%%%%%%%%%%%%%%%%%%%%%%%%%%%%%%%%%%%%%%%
%%%%%%%%%%%%%%%%%%%%%%%%%%%%%%%%%%%%%%%%%%%%%%%%%%%%%%%%%%%%%%%%%%%%%%%%%%%%%%%
{\em Summary}
%%%%%%%%%%%%%%%%%%%%%%%%%%%%%%%%%%%%%%%%%%%%%%%%%%%%%%%%%%%%%%%%%%%%%%%%%%%%%%%
%%%%%%%%%%%%%%%%%%%%%%%%%%%%%%%%%%%%%%%%%%%%%%%%%%%%%%%%%%%%%%%%%%%%%%%%%%%%%%%
%
As we have demonstrated above, it is possible to construct a quantum
simulator for the $O(3)$ nonlinear $\sigma$-model with present-day
technology. 
Such a restricted quantum computer would allow the comparison, for a 
controllable scenario, between perturbative and non-perturbative
analytical methods 
(renormalization flow \cite{asy-free,renormalize,cond-mat}, 
instantons \cite{instantons}, 
operator product expansion and vacuum condensates, 
low-energy theorems and sum rules \cite{general}, 
the S-matrix \cite{exact} etc.) 
as well as numerical results \cite{lattice} 
on the one hand with real quantum simulations on the other hand.
In contrast to most of the numerical simulations, for example, the
proposed quantum simulator works in real (laboratory) time, i.e.,  
it is not necessary to perform a Wick rotation to Euclidean time.
This advantage facilitates the study of the evolution of excitations,
for example collisions (S-matrix etc.).

Furthermore, the proposed set-up allows a direct access to the quantum
state and hence an investigation of the strong entanglement 
(e.g., in the ground state or near the quantum phase transition).
This could be done via state-selective radio/micro-wave spectroscopy
of transitions from the levels in Eq.~(\ref{ell}) to some higher-lying
empty and isolated internal level (of the semi-conductor) with a sharp
energy, for example (fluorescence measurement).  
Generating the radio/micro-waves via a circuit (wave-guide)
facilitates the position control of the measurement 
(vicinity of the inductance loop). 
Furthermore, one may also switch off the wires 
(e.g., by locally increasing the temperature) 
before the measurement.

It is also possible to create particles (and their anti-particles), 
which can be used to study the $S$-matrix, for example, via the
illumination with (left and right) circular polarized radio/micro-wave
radiation, cf.~Eq.~(\ref{charge}) and the subsequent remarks. 
Another interesting point is the robustness or fragility of 
non-perturbative properties (such as the instanton density) 
with respect to a small coupling to external degrees of freedom.

Apart from above points of interest, the construction of such a
restricted quantum computer, which is especially dedicated to the
simulation of the $O(3)$ nonlinear $\sigma$-model, would be an
interesting feasibility study for more general quantum simulators for 
a comparably well understood (yet nontrivial) system.
Finally, experience shows that the availability of a new tool 
(such as the proposed quantum simulator) yielding new tests/results  
usually leads us to a new level of understanding in physics with 
possibly unexpected outcomes.

%%%%%%%%%%%%%%%%%%%%%%%%%%%%%%%%%%%%%%%%%%%%%%%%%%%%%%%%%%%%%%%%%%%%%%%%%%%%%%%
%%%%%%%%%%%%%%%%%%%%%%%%%%%%%%%%%%%%%%%%%%%%%%%%%%%%%%%%%%%%%%%%%%%%%%%%%%%%%%%
{\em Outlook}
%%%%%%%%%%%%%%%%%%%%%%%%%%%%%%%%%%%%%%%%%%%%%%%%%%%%%%%%%%%%%%%%%%%%%%%%%%%%%%%
%%%%%%%%%%%%%%%%%%%%%%%%%%%%%%%%%%%%%%%%%%%%%%%%%%%%%%%%%%%%%%%%%%%%%%%%%%%%%%%
%
After having handled and understood the 1+1 dimensional situation, 
the extension to 2+1 dimensions should not be very problematic.
The 2+1 dimensional $O(3)$ nonlinear $\sigma$-model loses some of the
properties discussed above, but also acquires novel features, such as
skyrmions which are described by the topological current 
$j^\rho=\epsilon^{\mu\nu\rho}\,\f{\sigma}\cdot
(\partial_\mu\f{\sigma}\times\partial_\nu\f{\sigma})$.
The inclusion of an explicit $O(3)$-symmetry breaking term
$\f{n}\cdot\f{\sigma}$ should be easy in 1+1 and 2+1 dimensions.
Note that we did not incorporate a topological (Chern-Simons type) 
$\theta$-term  
${\cal L}_\theta=\theta\,\epsilon^{\mu\nu}\,\f{\sigma}\cdot
(\partial_\mu\f{\sigma}\times\partial_\nu\f{\sigma})$ in our 
1+1 dimensional scenario 
(in analogy to the $\theta$-term $G^*_{\mu\nu}G^{\mu\nu}$ in QCD), 
whose implementation is less straight-forward. 
Further interesting topics are the behavior of strongly interacting
quantum field theories (such as QCD and the $\sigma$-model) during the
cosmic expansion and the (long-range) entanglement of QCD vacuum
state (which might be used as a tool for diagnosis and a resource).

%
% SQUID ???
%

%%%%%%%%%%%%%%%%%%%%%%%%%%%%%%%%%%%%%%%%%%%%%%%%%%%%%%%%%%%%%%%%%%%%%%%%%%%%%%%
%%%%%%%%%%%%%%%%%%%%%%%%%%%%%%%%%%%%%%%%%%%%%%%%%%%%%%%%%%%%%%%%%%%%%%%%%%%%%%%
{\em Acknowledgments}\quad
%%%%%%%%%%%%%%%%%%%%%%%%%%%%%%%%%%%%%%%%%%%%%%%%%%%%%%%%%%%%%%%%%%%%%%%%%%%%%%%
%%%%%%%%%%%%%%%%%%%%%%%%%%%%%%%%%%%%%%%%%%%%%%%%%%%%%%%%%%%%%%%%%%%%%%%%%%%%%%%
%
R.~S.~gratefully acknowledges fruitful discussions with I.~Affleck,
M.~Krusius, P.~Stamp, B.~Unruh, G.~Volovik, and E.~Zhitnitsky
as well as support by the Humboldt foundation, the COSLAB
Programme of the ESF, CIAR, and NSERC. 
This work was supported by the Emmy-Noether Programme of the German
Research Foundation (DFG) under grant ${\rm No.~SCHU\;1557/1-1}$.

%%%%%%%%%%%%%%%%%%%%%%%%%%%%%%%%%%%%%%%%%%%%%%%%%%%%%%%%%%%%%%%%%%%%%%%%%%%%%%%
%%%%%%%%%%%%%%%%%%%%%%%%%%%%%%%%%%%%%%%%%%%%%%%%%%%%%%%%%%%%%%%%%%%%%%%%%%%%%%%
%%%%%%%%%%%%%%%%%%%%%%%%%%%%%%%%%%%%%%%%%%%%%%%%%%%%%%%%%%%%%%%%%%%%%%%%%%%%%%%
%%%%%%%%%%%%%%%%%%%%%%%%%%%%%%%%%%%%%%%%%%%%%%%%%%%%%%%%%%%%%%%%%%%%%%%%%%%%%%%

%%%%%%%%%%%%%%%%%%%%%%%%%%%%%%%%%%%%%%%%%%%%%%%%%%%%%%%%%%%%%%%%%%%%%%%%%%%%%%%
%%%%%%%%%%%%%%%%%%%%%%%%%%%%%%%%%%%%%%%%%%%%%%%%%%%%%%%%%%%%%%%%%%%%%%%%%%%%%%%
%%%%%%%%%%%%%%%%%%%%%%%%%%%%%%%%%%%%%%%%%%%%%%%%%%%%%%%%%%%%%%%%%%%%%%%%%%%%%%%
%%%%%%%%%%%%%%%%%%%%%%%%%%%%%%%%%%%%%%%%%%%%%%%%%%%%%%%%%%%%%%%%%%%%%%%%%%%%%%%
%%%%%%%%%%%%%%%%%%%%%%%%%%%%%%%%%%%%%%%%%%%%%%%%%%%%%%%%%%%%%%%%%%%%%%%%%%%%%%%
%%%%%%%%%%%%%%%%%%%%%%%%%%%%%%%%%%%%%%%%%%%%%%%%%%%%%%%%%%%%%%%%%%%%%%%%%%%%%%%
%%%%%%%%%%%%%%%%%%%%%%%%%%%%%%%%%%%%%%%%%%%%%%%%%%%%%%%%%%%%%%%%%%%%%%%%%%%%%%%
%%%%%%%%%%%%%%%%%%%%%%%%%%%%%%%%%%%%%%%%%%%%%%%%%%%%%%%%%%%%%%%%%%%%%%%%%%%%%%%
%%%%%%%%%%%%%%%%%%%%%%%%%%%%%%%%%%%%%%%%%%%%%%%%%%%%%%%%%%%%%%%%%%%%%%%%%%%%%%%
%%%%%%%%%%%%%%%%%%%%%%%%%%%%%%%%%%%%%%%%%%%%%%%%%%%%%%%%%%%%%%%%%%%%%%%%%%%%%%%
%%%%%%%%%%%%%%%%%%%%%%%%%%%%%%%%%%%%%%%%%%%%%%%%%%%%%%%%%%%%%%%%%%%%%%%%%%%%%%%
%%%%%%%%%%%%%%%%%%%%%%%%%%%%%%%%%%%%%%%%%%%%%%%%%%%%%%%%%%%%%%%%%%%%%%%%%%%%%%%
%%%%%%%%%%%%%%%%%%%%%%%%%%%%%%%%%%%%%%%%%%%%%%%%%%%%%%%%%%%%%%%%%%%%%%%%%%%%%%%
%%%%%%%%%%%%%%%%%%%%%%%%%%%%%%%%%%%%%%%%%%%%%%%%%%%%%%%%%%%%%%%%%%%%%%%%%%%%%%%
%%%%%%%%%%%%%%%%%%%%%%%%%%%%%%%%%%%%%%%%%%%%%%%%%%%%%%%%%%%%%%%%%%%%%%%%%%%%%%%
%%%%%%%%%%%%%%%%%%%%%%%%%%%%%%%%%%%%%%%%%%%%%%%%%%%%%%%%%%%%%%%%%%%%%%%%%%%%%%%
\end{document}